\UseRawInputEncoding
\documentclass[11pt]{article}
\usepackage{cite}
\usepackage{amsmath,amsfonts,amssymb}
\usepackage[small,bf,hang]{caption}
\usepackage{slashed}

\def\hybrid{
        \topmargin -20pt
        \oddsidemargin 0pt
        \headheight 0pt \headsep 0pt
        \textwidth 6.25in
        \textheight 9.5in
        \marginparwidth .875in
        \parskip 5pt plus 1pt \jot = 1.5ex}

\hybrid

 \csname
@addtoreset\endcsname{equation}{section}

\def\moth{\mathsurround=0pt}
\newdimen\zo \zo=0pt

\def\tick{\leaders\hrule height 0.5ex depth 0pt \hskip 0.5pt}
\def\upboxfill{$\moth \setbox\zo\hbox{\tick}%
  \hskip 3pt\hbox to 0pt{$\tick$\hss}\hrulefill \hbox to 7.5pt{$\tick$\hss}$}

\def\dtick{\leaders\hrule height .34pt depth 0.5ex \hskip 0.5pt}
\def\downboxfill{$\moth \setbox\zo\hbox{\dtick}%
  \hskip 2pt\hbox to 0pt{$\dtick$\hss}\hrulefill \hbox to 2pt{$\dtick$\hss}$}

\def\ciftS{\mathbb{S}}

\def\cH{{\cal H}}
\def\cK{{\cal K}}

\def\del{\partial}

\def\s{\sigma}
\def\n{\nu}
\def\a{\alpha}
\def\th{\theta}
\def\l{\lambda}

\def\be{\begin{equation}}
\def\ee{\end{equation}}
\def\bea{\begin{eqnarray}}
\def\eea{\end{eqnarray}}
\def\ba{\begin{array}}
\def\ea{\end{array}}

\begin{document}

\begin{titlepage}
\rightline{} \rightline\today
\begin{center}
\vskip 2.5cm {\Large \bf {  $Pin(d,d)$ covariance of pure spinor equations for  supersymmetric vacua and Non-Abelian T-duality}}\\
\vskip 1.2cm {\large {Aybike \c{C}atal-\"{O}zer} and Emine
Diri\"{o}z} \vskip 1cm
{\it {Department of Mathematics,}}\\
{\it {\.{I}stanbul Technical University,}}\\
{\it {Maslak 34469,
\.{I}stanbul, Turkey}}\\
ozerayb@itu.edu.tr \\ dirioz@itu.edu.tr \\

\vskip 1cm {\bf Abstract}
\end{center}

\vskip 0.5cm

\noindent
\begin{narrower}

\noindent In a supersymmetric compactification of Type II
supergravity, preservation of ${\cal{N}} = 1$ supersymmetry in
four dimensions requires that the structure group    of the
generalized tangent bundle $TM \oplus T^*M$ of the six dimensional
internal manifold $M$ is reduced from $SO(6,6)$ to $SU(3) \times
SU(3)$. This topological condition on the internal manifold
implies existence of two globally defined compatible pure spinors
$\Phi_1$ and $\Phi_2$ of non-vanishing norm. Furthermore, these
pure spinors should satisfy certain first order differential
equations. In this paper, we show that Non-Abelian T-duality
(NATD) is a solution generating transformation for these pure
spinor equations. We first show that the pure spinor equations are
covariant under $Pin(d,d)$ transformations. Then, we use the fact
NATD is generated by a coordinate dependent $Pin(d,d)$
transformation. The key point is that the flux produced by this
transformation is the same as the geometric flux associated with
the isometry group, with respect to which one implements  NATD. We
demonstrate our method by studying NATD of certain solutions of
Type IIB supergravity with $SU(2)$ isometry and $SU(3)$ structure.

\end{narrower}

\vspace{4cm}

\end{titlepage}

\newpage

\tableofcontents

\newpage

\section{Introduction}\label{introduction}

 Non-Abelian T-duality (NATD) is an extension of Abelian T-duality, which works
 well as a solution generating mechanism for string backgrounds with
 non-Abelian isometries. Although the rules for NATD for the  metric, the B-field and the dilaton
 field has been known for almost thirty years \cite{delaOssa:1992vci,Alvarez:1994np}, those for the RR fields has been understood relatively
 recently \cite{Sfetsos:2010uq}.
 Since then, NATD  has been widely used to generate new
 supergravity solutions with interesting holographic duals, see for example
\cite{Lozano:2011kb}-\cite{Itsios:2017cew}.

Recently, NATD has been described as a coordinate dependent
$O(d,d)$ transformation,
\cite{Hassler:2017yza,Lust,Demulder:2018lmj,ayb2,Sakatani,Bugden:2019vlj}.
The transformation under NATD of the metric, the B-field and the
dilaton field is determined by a coordinate dependent $O(d,d)$
matrix, which we will be calling $T_{\rm NATD}$, and the RR fields
transform under the corresponding $Pin(d,d)$ transformation
generated by $S_{\rm NATD}$. Here $\rho(S_{\rm NATD}) = T_{\rm
NATD}$, and $\rho$ is the usual double covering homomorphism from
$Pin(d,d)$ to $O(d,d)$. This approach makes it possible to view
NATD as a solution generating transformation  for Double Field
Theory (DFT), a framework which provides an $O(d,d)$ covariant
formulation for effective string actions
\cite{Tseytlin1,{dftRR},HullZ1,HullZ2} by introducing dual,
winding type coordinates. Since $T_{\rm NATD}$ is not constant, it
is not immediate that NATD should be a solution generating
transformation for DFT. However,  it is a special $O(d,d)$ matrix,
determined by the structure constants $C_{ij}^{\ k}$ of the
isometry algebra of the original background, and viewed as a twist
matrix within the framework of Gauged Double Field Theory, it
gives rise to geometric flux $ f_{ij}^{\ k} = C_{ij}^{\ k} $. This
is the key point in showing that NATD is a solution generating
transformation in DFT, which then provides a unified framework to
prove that it is a solution generating transformation for Type II
(generalised) supergravity. For details, see \cite{ayb2}. For a
similar approach, also see \cite{Sakatani}.

An important question is whether supersymmetry is preserved under
NATD. This problem is addressed in various papers, notably in
\cite{1409.7406,Zayas:2015azn,Itsios:2017cew,1408.6545,
1509.04286,1305.7229,1312.4945}. In \cite{1409.7406}, the
transformation under NATD of the gravitino and dilatino
supersymmetry variations were shown to be the same provided that
the Killing spinors did not depend on the isometry directions
along which NATD was applied. Equivalently, supersymmetry was
shown to be preserved (at least for a large class of backgrounds
with $SU(2)$ isometry) if the Killing spinors had vanishing
Kossmann-Lie derivative with respect to the Killing vector fields
generating the isometry. In the papers
\cite{1305.7229,1312.4945,1408.6545,1509.04286} NATD is applied to
backgrounds with ${\cal{N}} = 1$ supersymmetry. For such
backgrounds, conditions for supersymmetry can be described by
using tools from generalised geometry \cite{hitchin,{gualtieri}},
as was first shown in \cite{grana}. In this case, equations coming
from supersymmetry variations can be shown to be equivalent to a
set of differential equations to be obeyed by two globally defined
pure spinors. This will be discussed in detail in section
\ref{purespinorsection}. It is possible to apply NATD directly on
these pure spinors and check whether the transformed pure spinors
still satisfy the differential equations coming from
supersymmetry. This was the approach taken in
\cite{1305.7229,1312.4945,1408.6545,1509.04286}, where various
backgrounds with interesting holographic duals were examined. In
each case, the geometry supports an $SU(3)$ structure with
associated  pure spinors, and  it was checked by direct
computation that the NAT dual of these pure spinors indeed
satisfied the supersymmetry equations proving that NAT dual
background also preserved at least ${\cal{N}} = 1$ supersymmetry.
For a similar approach where one works with backgrounds supporting
a $G_2$ structure, see \cite{1402.3294}. It should be noted that
the Kossmann derivative of the Killing spinors along the isometry
directions vanish if and only if the Lie derivative of the pure
spinors (constructed as bilinears of these Killing spinors) along
these directions vanish. This condition was met by all  the
examples considered in the papers mentioned above.

In this paper,  we  describe the transformation of pure spinors
under NATD viewed as a $Pin(d,d)$ transformation by utilizing the
tools developed in \cite{ayb2}. This  enables us to prove that NAT
dual of pure spinors of ${\cal{N}} = 1$ vacua   still satisfy
these differential equations (and hence, the dual background will
also preserve at least ${\cal{N}} = 1$ supersymmetry since Bianchi
identities are also satisfied as shown in \cite{ayb2}), provided
that they have vanishing Lie derivative along the isometry
directions. To this end, we will first prove that the pure spinor
equations for preserved ${\cal{N}} = 1$ supersymmetry are
$Pin(d,d)$ covariant, by embedding them in DFT. Among other
things, using the framework of DFT makes it easier to show that
the  action of the exterior derivative operator and the $Pin(d,d)$
transformation must (anti-)commute, which is the trickiest part in
the proof. This is when the $Pin(d,d)$ matrix is constant. When it
is not constant, as is the case with NATD,  the pure spinor
equations will not be left invariant. However, NATD is generated
by a very special $Pin(d,d)$ transformation yielding geometric
flux, as we discussed above and again, this becomes the key point
in showing that solutions of pure spinor equations are still
solutions after NATD, provided that the pure spinors have
vanishing Lie derivative along the isometry directions. Compared
to the methods already present in the literature
our method has various advantages. First of all, describing the
dualisation of pure spinors as a $Pin(d,d)$ transformation makes
the computations rather direct, as it is not needed to specify an
ansatz for the seed background, as long as the isometry is
respected by the whole geometry, the fields and the pure spinors.
In particular, our proof is valid for any isometry group, not just
$SU(2)$\footnote{We will only discuss the case where the isometry
group acts without isotropy. If  not, our methods can be
generalised with some extra care.}. We should also note that our
method makes the determination of the $G$ structure of the NAT
dual background rather straightforward. In the particular examples
we will study in section \ref{examples}, the seed background will
be assumed to support $SU(3)$ structure, and we will see directly
how the associated pure spinors are transformed to pure spinors
associated with an $SU(2)$ structure. More generally, starting
with a pure spinor associated with a generic $SU(3) \times SU(3)$
structure, it is possible to work out the G-structure of the NAT
dual background, as is done in \cite{0807.4527} for Abelian
T-duality. In this paper, we will focus on the invariance of the
${\cal{N}}= 1$ supersymmetry equations on pure spinors under NATD
and will leave the discussion of the transformation of a generic
$SU(3) \times SU(3)$ structure to future study.

The plan of this paper is as follows: In Section \ref{review}, we
review the methods developed in \cite{ayb2}. In section
\ref{purespinorsection}, we focus on the pure spinor equations,
which were shown in \cite{grana, 2005,0609124} to be equivalent to
the supersymmetry equations to be satisfied by Type II vacua with
at least ${\cal{N}} = 1$ supersymmetry. We embed these equations
in the framework of Double Field Theory, so  that the covariance
under a general constant $Pin(d,d)$ transformation becomes
manifest. Then, in a separate subsection we discuss  the case when
the $Pin(d,d)$ transformation is coordinate dependent  (as it
happens for NATD), and show that whether the transformed pure
spinors satisfy the differential equations or not is completely
determined by the fluxes generated by the $Pin(d,d)$
transformation. Section \ref{examples} is devoted to explicit
examples. This is the section where we consider the ansatz for
Type IIB supergravity studied in \cite{1509.04286}. This ansatz is
general enough to cover many examples that are important in the
context of AdS/CFT duality. We transform the pure spinors
associated with the $SU(3)$ structure supported by the geometry by
applying the $Pin(d,d)$ transformation generating the NATD and
show that the resulting pure spinors (as well as the resulting
metric, B field and the RR fields) are in agreement with the ones
found in \cite{1509.04286}. We end with a discussion of results
and future directions in section \ref{conclusion}.

\section{Non-Abelian T-duality as an $O(d,d)$ transformation}\label{review}

The purpose of  this preliminary section  is to review  the
methods developed in \cite{ayb2}, where it was  shown that NATD of
a given $d$ dimensional Type II background with isometry $G$ can
be obtained through the action of a  coordinate dependent $O(d,d)$
matrix (also called $T_{{\rm NATD}}$) obtained by embedding the
following $O(n,n)$ matrix: \be \label{NATDmatrix} T_{{\rm NATD}} =
\left(\begin{array}{cc} 0 & 1_n \\
                        1_n & \theta_{ij} \end{array}\right), \ \ \ \
                        \theta_{ij}
                        = \nu_k C_{ij}^{\ \ k}. \ee
in $O(d,d)$ in the standard way (see 4.2.28-4.2.29 of
\cite{9401139}). Here, $\nu_k$ are coordinates of the NAT dual
background, and $C_{ij}^{\ \ k}$ are the structure constants of
the $n$ dimensional Lie algebra of the isometry group $G$, so $i,
j, k= 1, \dots, n.$ The presence of the Lie group $O(d,d)$, which
is the global symmetry group of DFT, makes it possible to describe
the transformation under NATD of the Type II supergravity fields
as a transformation in DFT. More precisely, one rewrites the
supergravity fields in terms of the DFT fields $\cH, d, \chi$,
where $\cH$ is the generalized metric that encodes the metric and
the B-field, $d$ is the generalized dilaton field and $\chi$ is
the spinor field that packages the modified RR fields of Type II
supergravity in the democratic formulation. These fields, being
solutions of Type II supergravity also solve the DFT equations in
the supergravity frame\footnote{DFT is consistent only when one
imposes the so called strong constraint, that effectively
eliminates half of the doubled coordinates. This constraint is
trivially satisfied when the DFT fields and gauge parameters are
independent of the winding type coordinates. In this case, the DFT
fields are said to belong to the supergravity frame, since the DFT
action and field equations reduce to those of Type II supergravity
in the democratic formulation.}. As it is assumed that the
isometry is respected by all the fields in the background, it is
possible to go to a non-holonomic frame so that the DFT fields,
when written with respect to such a frame, are independent of the
isometry coordinates. In \cite{ayb2} we refer to such fields as
\emph{untwisted fields}. Plugging the initial DFT fields in the
field equations of DFT (of both the NS-NS sector and RR sector of
Type II supergravity), one sees that the untwisted DFT fields
satisfy the field equations of Gauged Double Field Theory
(GDFT)\footnote{GDFT is a deformation of DFT, obtained from a
Scherk-Schwarz reduction and the deformation is determined
entirely by the  fluxes associated with the Scherk-Schwarz twist
matrix
\cite{Grana:2012rr,Geissbuhler:2011mx,Aldazabal:2011nj,ozer}.},
with geometric fluxes associated with isometry. It was shown in
\cite{ayb2} that the NAT dual DFT fields $\cH', d', \chi'$ are
found by acting on the untwisted fields $\cH(x), d(x), \chi(x)$
\footnote{Here, we call the spectator coordinates excluding the
isometry directions collectively $x$ and the doubled coordinates
of the NAT dual background collectively $\nu$.} by the $O(d,d)$
matrix (\ref{NATDmatrix}) as below:
\begin{eqnarray} \label{NATDH} \cH'(x, \nu) &=& T_{{\rm
NATD}}(\nu)
\cH(x) (T_{{\rm NATD}})^t(\nu)\\
\label{NATDS} \cK'(x,  \nu ) &=& S_{{\rm NATD}}(\nu)
\cK(x) (S_{{\rm NATD}})^{-1}(\nu)\\
\label{NATDF} F'(x, \nu) &=& e^{-\sigma(\nu)} e^{-B'(x, \nu)}
S_{{\rm NATD}}(\nu)
e^{B(x)} F(x) \\
\label{NATDd} d'(x, \nu) &=& d(x) + \sigma(\nu).
\end{eqnarray} Here, $\cK = C_d^{-1} \ciftS$, and $C_d$ is given in (\ref{chargeconj}) in Appendix
\ref{appC}. $\ciftS$ is the element in $Spin^-(d,d)$ that projects
onto $\cH$ under the double covering homomorphism $\rho$ between
$Pin(d,d)$ and $O(d,d)$, that is $\rho(\ciftS) = \cH$. Similarly,
$\rho(S_{{\rm NATD}}) = T_{{\rm NATD}}$ and up to a sign it is
given as \cite{ayb2}
 \be \label{SNATDmatrix} S_{{\rm NATD}}= C_n S_{\theta} = S_{\beta}
C_n. \ee  The factors $S_{\theta}$ and $S_{\beta}$ in $S_{{\rm
NATD}}$ are the $Spin^+(10,10)$ elements that projects onto the
$SO^+(10,10)$ matrix that generates the $B$-transformations and
$\beta$-shifts with $\theta_{ij}
                        = \nu_k C_{ij}^{\ \ k}$ and $\beta_{ij}
                        = \nu_k C_{ij}^{\ \ k}$, respectively.
$B'(x, \nu) $ that appears in (\ref{NATDF}) is read off from
$\cH'(x, \nu)$ in (\ref{NATDH}). The field $\sigma(\nu)$ in
(\ref{NATDd}) and (\ref{NATDF}) is non-vanishing only when the
isometry group is non-unimodular. For the purposes of this paper,
it can be taken to be zero. The fact that the NAT dual fields can
be written in terms of DFT fields as in (\ref{NATDH}-\ref{NATDd})
makes it straightforward to prove that NATD is a solution
generating transformation for the field equations of Type II
supergravity. In fact, all one has to do is to show that the
fields in (\ref{NATDH}-\ref{NATDd}) solve the DFT equations, since
the coordinates $(x, \nu)$ can be identified with the physical
space-time coordinates, putting all the fields in the supergravity
frame. Due to the special form of the fields, this amounts to
showing that the untwisted fields $\cH(x), d(x), F(x)$ appearing
on the right hand side of (\ref{NATDH}-\ref{NATDd}) solve the
field equations of GDFT, with fluxes generated by $T_{{\rm
NATD}}$. Now, the key point is that this is  exactly the same as
the geometric flux associated with the isometry group, that is, $
f_{ij}^{\ k} = C_{ij}^{\ k} $, and we already know that the
untwisted fields satisfy the GDFT equations with geometric flux.
As a result, one concludes that NATD is a solution generating
transformation for Type II supergravity, both in the NS-NS and the
RR sector, simply owing to the fact that fluxes are preserved. The
idea that preservation of flux should be a guiding principle in
determining whether an $O(d,d)$ transformation is a solution
generating transformation for supergravity has also been used in
\cite{Borsato:2020bqo,Borsato2,Hassler:2020tvz,Borsato3} and very
recently in \cite{Borsato:2021vfy}. A similar approach was applied
in \cite{Gubarev} to find solution generating U-duality
transformations within the framework of exceptional field theory.
In the next section, we will see that the same principle  also
plays a key role in examining preservation of supersymmetry under
NATD.

\section{Covariance of Pure Spinor Equations under $Pin(d,d)$}\label{purespinorsection}

As was first shown in the seminal paper \cite{2005}, the
conditions to be obeyed by the internal space in a supersymmetric
compactification of Type II supergravity can be neatly described
within the framework of generalized complex geometry
\cite{hitchin, gualtieri}. Demanding that the four dimensional
solution preserves at least  ${\cal{N}} = 1$ supersymmetry implies
that the structure group of the generalized tangent bundle $TM
\oplus T^*M$ of the six dimensional internal manifold $M$ is
reduced from $SO(6,6)$ to $SU(3) \times SU(3)$. This topological
condition on the internal manifold implies the existence of two
globally defined compatible pure spinors $\Phi_1$ and $\Phi_2$ of
non-vanishing norm. These $Cliff(6,6)$ spinors can be constructed
from the internal spinors arising from the 10 dimensonal Killing
spinors generating the supersymmetry transformations in 10
dimensions. A $Cliff(6,6)$ spinor can be mapped to a
non-homogenous differential form (a polyform) through the Clifford
map. It was shown in \cite{grana,2005,0609124} (also see
\cite{0507099}) that the Killing spinor equations coming from
supersymmetry variations is equivalent to the following
differential equations for the two pure spinors: \bea  d(e^{2A - \phi} e^{B} \wedge \Phi_1) & = & 0 , \label{pure1} \\
 d(e^{2A - \phi} e^{B} \wedge \Phi_2) & = & e^{2A-\phi} dA \wedge e^{B} \wedge \bar{\Phi}_2  + \frac{i}{8} e^{3A}
e^{B} \wedge \lambda(*_6 F).  \label{pure2}
 \eea
For computational details on derivation of these equations, see
Appendix A of \cite{0609124} and Appendix B of \cite{0507099}. For
the corresponding equations for general ten dimensional
supersymmetric solutions which do not necessarily involve a four
dimensional Minkowski space factor, see \cite{Tomasiello:2011eb}
(the equations above  are discussed as a special case in their
section 4.1). Note that for our purposes, we have presented the
equations in a form where the B field appears explicitly, rather
than writing them in terms of the differential operator $d_H = d +
H \wedge$ as was originally done in
\cite{grana,2005,0609124}.\footnote{\label{Bconvention} In fact,
it is more common in the literature to express these equation in
terms of the operator $d_H = d - H \wedge$. This involves a field
redefinition $H \rightarrow -H$ for Type IIA with respect to the
conventions of \cite{Bergshoeff:2001pv}. In section \ref{examples}
we will be looking at a IIA background with non-trivial B-field,
so we prefer to agree with the conventions of
\cite{Bergshoeff:2001pv} for Type IIA (since the conventions
adopted in \cite{ayb2} agree with those of
\cite{Bergshoeff:2001pv}), and this means we need the above field
redefinition for Type IIB. This also means that our convention for
the B-field is the opposite of that of \cite{1509.04286}. Indeed,
the B-field we find in \eqref{newgB} in section \ref{examples} has
opposite sign compared to the B-field found in \cite{1509.04286}.}

In the equations above, $A$ is the warp factor that appears in the
compactification ansatz \be \label{metric} ds^2 = e^{2A(y)}
dx^2_{3,1} + g_{mn} dy^m dy^n,
\ \ m, n = 1, \cdots, 6. \ee
$\phi$ is the dilaton field and $*_6$ is the Hodge duality on the
six dimensional internal manifold. $F$ is related to the polyform
$F^{(10)}$ that encodes the RR fields in the democratic
formulation of supergravity \cite{Bergshoeff:2001pv} in the
following way \be \label{forms} F^{(10)} = F + \hbox{vol}_4 \wedge
*_6 (\lambda F). \ee Here, $F = F_0 + F_2 + F_4 + F_6$   for Type
IIA and  $F = F_1 + F_3 + F_5$ for Type IIB, and they are internal
forms having components only along the six dimensional internal
space. Also, \be \label{lambda} \lambda(A_n) \equiv
(-1)^{Int[n/2]}A_n = (-1)^{n(n-1)/2} A_n\ee for an $n$-form $A_n$.
As a $Spin(d,d)$ spinor $F$ has positive chirality for Type IIA
and is of negative chirality for Type IIB. The chirality of the
pure spinor $\Phi_1 $ is the same as that of the RR fluxes and the
pure spinor $\Phi_2$ has opposite chirality.

In the next two subsections, we will  show that these equations
are covariant under $Pin(d,d)$ transformations.

\subsection{Constant $Pin(d,d)$ transformation}\label{s1}

Although (\ref{NATDmatrix}) is non-constant,  we start by
considering the transformation of the pure spinor equations under
a constant $O(d,d)$ matrix $T$ and the corresponding $Pin(d,d)$
matrix $P$ with $\rho(P) = T$, where $\rho$ is the
                           double covering homomorphism $$\rho :
                           Pin(d,d) \rightarrow O(d,d).$$
The transformation of the
 RR fluxes $F$ under $P \in Pin(d,d)$ is \cite{Fukuma} \be   \label{transF} F \rightarrow  F'
= P . F = e^{-B'} P e^{B} F. \ee Here, the transformation of the
B-field is read off from the antisymmetric part of the transformed
background matrix $ E \equiv g + B$: \be \label{transE} E
\rightarrow E'(g', B') = T . E(g, B) = (a E + b)(c E + d)^{-1} \ee
when  the $O(d,d)$ matrix $T$ is of the form \be T =
\left(\begin{array}{cc} a & b \\
                        c & d \end{array}\right). \ee Note that
                        this is equivalent to the aforementioned transformation of the generalized metric
                        \cite{zwiebachL}
\be \cH'(g', B') = T \cH(g,B) T^t. \ee

It is known that the pure spinors transform  under $Pin(d,d)$ in
basically the same way as the RR fields transform. However, there
is a slight change which makes sure that the norms of the pure
spinors are kept invariant (up to a sign). The norm $\parallel
\Phi
\parallel$ of a pure spinor $\Phi$ is defined \cite{grana,2005,0609124} via the Mukai pairing $< ,
>$, which is an invariant bilinear form on spinors (see Appendix \ref{appC}): \be \label{norm} < \Phi,
\bar{\Phi} > = -i \parallel \Phi \parallel^2 vol \ee where $vol$
is the volume form determined by the metric. As discussed in
detail in Appendix \ref{appC}, the Mukai pairing has the following
transformation property under the action of certain elements $P$
of $Pin(d,d)$: \be <P \Phi_1, P  \Phi_2
> = \pm < \Phi_1, \Phi_2 >, \ee where either $P \in Spin(d,d)$ or
is of the form $P = C_n S$ or $P = S C_n$ with $S \in Spin(d,d)$
and $C_n$ is as in (\ref{chargeconj}) (Recall that the NATD matrix
(\ref{SNATDmatrix}) is of this form). On the other hand the volume
form $vol = \star_d 1$ transforms as \be \nonumber \star_d 1 =
\sqrt{det g} dy^1 \wedge \cdots dy^d \rightarrow \star_d' 1 =
\sqrt{det g'} dy^1 \wedge \cdots dy^d = G \sqrt{det g} dy^1 \wedge
\cdots dy^d = G \star_d 1, \ee where
 \be G \equiv\frac{\sqrt{det g'}}{\sqrt{det
g}} = det(cE + d)^{-1}.\ee This follows immediately from the
transformation of the metric $g$ which can be read off from the
symmetric part of $E'$ in (\ref{transE}). Hence, the
transformation of the pure spinors under $Spin(d,d)$ must be
accompanied by a scale transformation with a factor of $\sqrt{G}$
\ \footnote{In the framework of generalized geometry, the pure
spinors $\Phi$ and $\sqrt{G} \Phi$ correspond to the same
generalized complex structure, as they belong to the same pure
spinor line sub-bundle of $\bigwedge^\bullet T^*$.}: \be
\label{transphi} \Phi  \rightarrow \Phi' = \sqrt{G} \ P . \Phi =
\sqrt{G} \ e^{-B'} P e^{B} \Phi \ee so that the norm (\ref{norm})
remains invariant up to a sign. Note that this extra factor of $
\sqrt{G}$ also ensures that \be \label{covariance1} e^{2A' -
\phi'} e^{B'} \wedge \Phi'_{1,2} = P \left(e^{2A - \phi} e^{B}
\wedge
  \Phi_{1,2} \right) \ee since $A$ is invariant and the dilaton
  field $\phi$ transforms exactly with the  same $\sqrt{G}$
  factor: \be \label{transdilaton} e^{ \phi'} = \sqrt{G} e^{\phi}.\ee
The transformation rule (\ref{transdilaton}) follows
  from the fact that the generalized dilaton field $ e^{-2d} = \sqrt{{\rm det} g } \ e^{-2 \phi}$ is invariant
  under $O(d,d)$ (consider the equation (\ref{NATDd}) with $\sigma = 0$), that is $e^{-2 d'} = e^{-2 d}$ so that:\be \label{dilaton1} e^{-2 \phi'} \sqrt{{\rm det }g'} = e^{-2
  \phi} \sqrt{{\rm det }g}. \ee

Now all we have to do is to figure out the transformation of the
term involving Hodge duality on the right hand side of equation
(\ref{pure2})  and also to show that the action of $P$ and the
exterior derivative operator $d$ on the $Clif(d,d)$ spinors
$\Phi_{1,2}$ and $F$ commutes.

 For both purposes, we find it useful to embed these
equations in Double Field Theory. Towards this we extend the
exterior derivative operator $d = \frac{1}{2}
 \Gamma^i \del_i$ to \be \label{der} d + \tilde{d} \equiv \frac{1}{2} \Gamma^M
 \del_M = \frac{1}{2}(\Gamma^i \del_i + \Gamma_i \tilde{\del}^i) = \psi^i \del_i + \psi_i \tilde{\del}^i. \ee Here, the gamma matrices
 $\Gamma^M = (\Gamma_i, \Gamma^i)$ are the  the Clifford algebra
elements satisfying the following Clifford product relations:
\be \label{clifford}
  \{ \Gamma_i ,  \Gamma^j\} \ = \ 2 \delta_{i}{}^{j}\;, \qquad
  \{\Gamma_i,\Gamma_j\} \ = \  0\;, \qquad \{\Gamma^i,\Gamma^j\} \ = \ 0\;,
 \ee For future reference we also defined in (\ref{der})  \be \label{psi} \psi^M \equiv \frac{1}{\sqrt{2}}
\Gamma^M.\ee Also, we
 write $\star \lambda(F)$ as \cite{dftRR,ozer}  \be \label{hodge} \star \lambda(F) = -C_d^{-1} S_g^{-1} F, \ee
 where $S_g^{-1} = S_{g^{-1}}$ is the $Spin(d,d)$ element that
 projects onto the $SO(d,d)$ element \be h_{g^{-1}} \equiv
 \left(\begin{array}{cc} g^{-1} & 0 \\
                           0 & g \end{array}\right) \ee under the
                           double covering homomorphism $\rho$ that is,
                           $\rho( S_{g^{-1}}) =  h_{g^{-1}}.$ Note
                           that the equation (\ref{hodge}) is valid
                           in all even dimensions\footnote{In odd dimensions, the definition of $\cK$ involves
                           $(\psi^i + \psi_i)$, rather than the $(\psi^i - \psi_i)$ in (\ref{chargeconj}). See \cite{dftRR} for
                           more details.}, in particular for $*_6$ with $d
                           = 6$.

 It is useful to write $C_d S_{g^{-1}}$ as $e^{-B} \cK_d e^{B}$ where
 $\cK_d = C_d^{-1} \ciftS$ and $\ciftS \equiv S_B^{\dagger} S_{g^{-1}} S_B$ is the $Spin(d,d)$ element that
 projects onto the generalized metric $\cH $. Indeed, \bea  e^{-B} \cK_d e^{B} & = & e^{-B} C_d^{-1} \ciftS e^{B} = e^{-B} C_d^{-1}
S_B^\dagger S_{g^{-1}} S_B
 e^{B}\\
 & = & e^{-B} C_d^{-1} C_d e^{B} C_d^{-1} S_{g^{-1}} e^{-B} e^{B} = C_d^{-1}
 S_{g^{-1}}. \eea where we have used that $S_B = e^{-B}$ and
 $S_B^\dagger = C_d S_{-B} C_d^{-1} = C_d e^{B} C_d^{-1}.$
 Rewriting (\ref{hodge}) for $d=6$ and in terms of $\cK_d$, we have
 \be \label{hodgesix} *_6 \lambda(F) = - e^{-B} \cK_6 e^{B} F. \ee Rewriting the equations
 (\ref{pure1}) and (\ref{pure2}) we get
 \bea  \Gamma^M \del_M(e^{2A - \phi} e^{B} \wedge \Phi_1) & = & 0 , \label{covpure1} \\
 \Gamma^M \del_M(e^{2A - \phi} e^{B} \wedge \Phi_2) & = & e^{2A-\phi} \Gamma^M \del_M A \wedge e^{B} \wedge \bar{\Phi}_2  \mp \frac{i}{8} e^{3A}
\cK_6 e^{B} F.  \label{covpure2}
 \eea These equations reduce to equations (\ref{pure1}) and
 (\ref{pure2}) in the supergravity frame where fields do not depend
 on the winding type coordinates so that $\tilde{\del}^i = 0$. The
 upper sign in the last term of (\ref{covpure2}) is for Type IIB
 and the lower sign is for Type IIA. This is because in six
 dimensions $*_6 \lambda = \lambda *_6$ for odd degree forms,
 whereas $*_6 \lambda = - \lambda *_6$ for even degree forms.

We know that $F^{(10)}$ in (\ref{forms}) transforms as in
(\ref{transF}). Let us discuss what this implies for the
transformation of the internal forms $F$. We have \bea \nonumber
F^{(10)'} &=& e^{-B'} P e^{B} (F - {\rm vol}_4 \wedge e^{-B} \cK_6
e^{B} F) \\
&=& e^{-B'} P e^{B} F - {\rm vol}_4 \wedge e^{-B'} P \cK_6 e^{B}
F. \label{decomp} \eea where we have used (\ref{hodgesix}) and the
fact that ${\rm vol}_4 $, being an even form, commutes with all
elements of $Pin(d,d)$. To rewrite (\ref{decomp}) in the form
(\ref{forms}) we first define \be \label{newintform} F' \equiv
e^{-B'} P e^{B} F, \ee which is again an internal form, as all the
$Pin(d,d)$ operators on the left hand side have actions only on
the internal space and then use the fact that under $P \in
 Pin(d,d)$ the field $\cK_d$ transforms as \be \label{transK} \cK_d \rightarrow P .
 \cK_d = \cK'_d = P \cK_d P^{-1}. \ee Inserting a $P^{-1} P$ after $\cK$ in the
 second term of the right hand side of
 (\ref{decomp}) and using (\ref{hodgesix}) and (\ref{newintform}), we
 obtain \bea \nonumber F^{(10)'} &=& F' - {\rm vol}_4
\wedge e^{-B'} \cK'_6 e^{B'} F' \\
&=& F' + {\rm vol}_4 \wedge \tilde{*}_6 \lambda(F'). \eea Note
that $F'$ has components only along the six dimensional deformed
space and the Hodge duality is taken with respect to the metric
after the $O(d,d)$ transformation. This shows us that not only the
polyform $F^{(10)}$ that encodes the RR fields in the democratic
formulation, but also the internal polyform $F$ that appears in
the pure spinor equations (\ref{pure1},\ref{pure2}) transform in
the expected way as given in (\ref{newintform}).

Using the transformation properties (\ref{transK}) and
(\ref{newintform},\ref{transphi},\ref{transdilaton}) and  the fact
that $A$ is invariant under $Pin(d,d)$ we also see that
\be \label{covariance2}  e^{3A'} \cK_d' e^{B'} F' = P \left(e^{3A}
\cK_d e^{B} F \right). \ee In order to prove the covariance of the
pure spinor equations under $Pin(d,d)$ we next discuss whether or
not the generalized exterior derivative operator $\Gamma^M \del_M$
commutes with the action of $Pin(d,d)$. We first start with
$Spin(d,d)$ and show  \be \label{commute} \Gamma^M \del_M (S \
\chi) = S (\Gamma^M \del_M \chi), \ \ \ S \in Spin(d,d) \ee for
any spinor field $\chi$. Using the relations \be
\label{mainwithgamma} (h^{-1})^M_{\ A} \Gamma^A = S^{-1} \Gamma^M
S, \ee where $h$ is the $SO(d,d)$ element that satisfies
$\rho(S^{-1}) = h$, we see that for constant $S \in Spin(d,d)$:
\be \Gamma^M \del_M ( S  \ \chi) = \Gamma^M S \ \del_M \chi = S
\Gamma^A (h^{-1})^M_{\ A} \del_M \chi. \ee Then, the commutation
relation (\ref{commute}) holds, if we have  \be \label{cond1}
(h^{-1})^M_{\ A} \del_M \chi = \del_A \chi \ee Note that we would
have in DFT, \be (h^{-1})^M_{\ A} \del_M \chi(h X) = \del'_A
\chi(X'). \ee since one also transforms $ X \rightarrow X'= h X$.
However in all the examples we will be looking at, the
transformation generated by $P$ will act only along the
coordinates on which the pure spinors will not depend, so that we
will always have $ X ' = X $ and hence $\del'_A \chi = \del_A
\chi$. For example, if the background possesses $d$ commuting
isometries, it is possible to choose coordinates such that  the
fields depend on only $10-d$ of the 10 coordinates. Associated
with the $d$ isometries, there is an $O(d,d)$ Abelian T-duality
group acting on the background along these $d$ coordinates. Since
the coordinates have been chosen in such a way that none of the
fields (including the global spinor fields) do not depend on these
directions, we have $\del'_A \varphi(X) = \del_A \varphi(X)$,
where $\varphi$ denotes any field or gauge parameter in the theory
and $A$ runs through the $10-d$ coordinates. To summarize,
equation (\ref{commute}) holds as desired, as long as the
condition (\ref{cond1}) is satisfied. This immediately implies
(using
(\ref{transphi},\ref{transdilaton},\ref{transK},\ref{newintform})
and the invariance of $A$) that the pure spinor equations
(\ref{covpure1}, \ref{covpure2}) are covariant under $Spin(d,d)$
transformations. Note that there is no sign flip  in front of the
last term on the right hand side of (\ref{covpure2}), since
$Spin(d,d)$ transformations takes a solution of Type IIA/IIB to a
solution also of Type IIA/IIB. However, a $Pin(d,d)$
transformation which involves odd number of reflections maps a
solution of Type IIA to a solution of Type IIB and vice versa, and
hence the sign of the aforementioned term in (\ref{covpure2})
flips after the transformation. Despite this, the pure spinor
equations (\ref{covpure1}, \ref{covpure2}) are still covariant,
since for such $P$, the differential operator $d = \Gamma^M
\del_M$ and $P$ anti-commutes, as we will now discuss.

Consider the $Pin(d,d)$ elements $\Lambda_i$ given in
(\ref{lambda}). From the Clifford commutation relations
(\ref{clifford}) one can easily compute\footnote{Note that
(\ref{anticomm}) implies that $\rho(\Lambda_i) = h_i,$ where \be \label{hi} h_i = - \left(\ba{cc} 1 - E_i &  E_i \\
                                     E_i  & 1 - E_i \ea \right), \ \ (E_i)_{jk} = \delta_{ij}\delta_{ik}. \ee}  \be \label{anticomm}
 \Lambda_i . \Gamma^M . (\Lambda_i)^{-1}
                    =\left\{
  \begin{array}{l l l}
   -\Gamma_i & \quad \text{if\; $\Gamma^M = \Gamma^i$}\\
   -\Gamma^i & \quad \text{if\, $\Gamma^M = \Gamma_i$\,.}\\
   - \Gamma^M & \quad \text{otherwise}
  \end{array} \right.
                    \ee

In this paper, we will be looking at the $Pin(d,d)$ elements that
can be written as a product of $Spin(d,d)$ elements and
$\Lambda_i$, simply because the NATD matrix is of this form. Our
discussions here can be straightforwardly extended so as to
include the $Pin(d,d)$ elements which also involve the elements
$\Lambda^+_i$ given in (\ref{lambdaplus}), but we refrain from
doing that in order to avoid equations cluttered with pluses  and
minuses.

Due to the  relations (\ref{anticomm}), we see that \be
d(\Lambda_i \chi) = \Gamma^M \del_M (\Lambda_i \chi) = - \Lambda_i
\Gamma^M \del_M \chi = - \Lambda_i d\chi, \ee provided that $^M
\neq \ ^i$ or $^M \neq \ _i$, which then implies that the
differential $d = \Gamma^M \del_M$ and $P$ commutes if $P$
involves an even number of $\Lambda_i$s
 and they anti-commute
otherwise. As discussed above, this condition is automatically
satisfied for Abelian T-duality, due to the existence of $d$
commuting isometries. This makes it possible to choose a
coordinate system such that none of the fields depend on the
coordinates along which the (constant) $O(d,d) / Pin(d,d)$
transformation acts, and hence the desired commutation  or
anti-commutation relations hold. Therefore, we conclude that the
pure spinor equations are covariant under Abelian T-duality. As
for NATD, (\ref{cond1}) is also satisfied with a convenient choice
of coordinates (again due to existence of isometries), but we
still need to discuss the situation with non-constant $P$, since
the NATD matrix (\ref{NATDmatrix}) is not constant as has been
assumed above. This discussion will be carried out in the next
section.

Note that the covariance of the equations
(\ref{pure1},\ref{pure2}) for certain cases has been discussed
before, albeit in a different language. For example, in
\cite{0807.4527} the covariance of the pure spinor equations for
backgrounds with $U(1)$ isometry was shown.\footnote{More
precisely, they studied the factorized duality for $d=1$. See
\cite{9401139} for the discussion of how factorized duality,
B-shifts and $GL(d)$ transformations are embedded in the T-duality
group $O(d,d)$ for flat and curved backgrounds with $d$ commuting
isometries.} Another example is the Lunin-Maldacena (LM)
transformation (also called TsT transformation) which can be
described as on $O(2,2)$ transformation
\cite{LM,Frolov,CatalOzer}. In \cite{0606257} the transformation
of the pure spinors corresponding to an $SU(3)$ structure under
this $O(2,2)$ transformation was discussed within the framework of
generalized complex geometry, as we do here. That the transformed
pure spinors (now corresponding to an $SU(2)$ structure) still
satisfy the supersymmetry equations was also checked for this
particular $O(2,2)$ transformation.

\subsection{Non-constant $Pin(d,d)$ transformation}\label{s2}

In this subsection, we extend the discussion in the previous
subsection to the case where the $Pin(d,d)$ transformation (and
hence the corresponding $O(d,d)$ transformation) depends on some
of the internal coordinates. This is important, as the NATD
transformation and the Yang-Baxter transformation are known to be
generated by such coordinate dependent $Pin(d,d)$ transformations.
The transformation properties summarized in
(\ref{covariance1},\ref{covariance2}) are obviously still valid,
even when $P \in Pin(d,d)$ is coordinate dependent. However, one
has to be more careful in discussing the commutation of the
exterior derivative operator $d$ and the action of $P$, as now $d$
also acts on $P$.

Let us   first discuss the case when the $Pin(d,d)$ matrix does in
fact lie in the subgroup $Spin^+(d,d)$, $P = S  \in Spin^+(d,d)$
(so that we can use the useful identity (\ref{isom})): \bea
\label{commute2}
\Gamma^M \del_M (S \ \chi(X)) & = &   \left\{ \Gamma^M S \ \del_M  + \Gamma^M S (S^{-1} \partial_M S)\right\} \chi(X)  \\
& = &  S \ \Gamma^A  \ (h^{-1})^M_{\ A} \left( \del_M +
 S^{-1} \del_M S \right) \chi(X), \nonumber \eea
where $\rho(S^{-1}) = h$ and in passing to the second line, we
have used (\ref{mainwithgamma}). To calculate the second term in
(\ref{commute2}) we use  an important identity that follows from
the fact that the Lie algebras of $SO(d,d)$ and $Spin(d,d)$ are
isomorphic: \bea \label{isom} \Gamma^A \ (h^{-1})^M_{ \ \ A}
S^{-1}
\partial_M \ S &=& \frac{1}{4} \Omega_{ABC} \Gamma^A \ \Gamma^B \
\Gamma^C \nonumber \\
&=&  \frac{1}{12} f_{ABC} \Gamma^A \ \Gamma^B \ \Gamma^C \ \chi(X)
-\frac{1}{2} f_B \Gamma^B \ \chi(X). \eea Here, $f_{ABC}$ are the
fluxes associated with the matrix $S$ (see \cite{ozer,ayb2} for
the definition).

Now, we again assume that the transformation matrix $S$ is such
that (\ref{cond1}) is obeyed. We emphasize again that this
condition is trivially satisfied if the field $\chi$ does not
depend on the coordinates along which $S$ and hence $h$ acts
nontrivially. This is indeed the case for NATD and is guaranteed
by the fact that NATD acts along isometry directions. Then, under
this assumption, we have \be \label{gammaS} \Gamma^M \del_M (S
\chi) = S (\Gamma^A \nabla_A \chi), \ee where \be \label{nabla}
\nabla_A = \del_A + \frac{1}{12} f_{ABC}
 \Gamma^B \ \Gamma^C  -\frac{1}{2} f_A. \ee

Let us now discuss what happens when $P$ involves odd number of
$\Lambda_i$ factors, so that $P$ does not lie in the $Spin(d,d)$
subgroup (if the number of $\Lambda_i$  factors is even, then $S$
is still in $Spin(d,d)$, albeit not in the subgroup $Spin^+(d,d)$
connected to identity). For simplicity, we assume that $P$ is of
the form $P = C_n S$, where $S \in Spin^+(d,d)$ and $C_n$ is as in
(\ref{chargeconj}) with $n$ odd. Equation (\ref{mainwithgamma}) is
valid for all $Pin(d,d)$ elements, so we have \be
\label{mainwithgamma2} (h_1 \cdots h_n U)^M_{\ A} \Gamma^A =
P^{-1} \Gamma^M P, \ee where $U$ is the $SO^+(d,d)$ element that
satisfies $\rho(S) = U$, and $h_i$ satisfy $\rho(h_i) = \Lambda_i$
and are given in (\ref{hi}). When $n$ is odd, it can be easily
seen that $h_1 \cdots h_n = -J^d_n$, where $J^d_n$ is the $O(d,d)$
matrix obtained by embedding the $O(n,n)$ matrix  \be \label{jn}
J_n =
\left(\begin{array}{cc} 0 & 1_n \\
1_n & 0 \end{array}\right),\ee in $O(d,d)$ in the usual way (see
4.2.28-4.2.29 of \cite{9401139}). Rewriting the first line of
(\ref{commute2}) for $P = C_n S$ and using (\ref{mainwithgamma2})
we have \bea \label{commute3}
\Gamma^M \del_M (P \ \chi(X)) & = &   \left\{ \Gamma^M P \ \del_M  + \Gamma^M P (P^{-1} \partial_M S)\right\} \chi(X)  \\
& = & - P \ \Gamma^A  \ (J^d_n U)^M_{\ A} \left( \del_M +
 S^{-1} \del_M S \right) \chi(X), \nonumber \eea where we have
 also used $P^{-1} \del_M P = S^{-1} \del_M S$ for $P = C_n S$.
 Using (\ref{isom}) again, one can see that we have \be \label{gammaP} \Gamma^M \del_M (P  \chi) =
 -P (\Gamma^A \nabla_A \chi), \ee where $\nabla$ is as in
 (\ref{nabla}), now with fluxes $f'_{ABC} = (J^d_n)^D_{\ A} f_{DBC}$
 with $f$ being the fluxes associated with the $Spin^+(d,d)$
 matrix $S$.

Collecting the results in
(\ref{covariance1},\ref{covariance2},\ref{commute}, \ref{gammaS})
and (\ref{gammaP}), we conclude that the fields
 \emph{after} the transformation generated by the non-constant $P \in Pin(d,d)$  satisfy the supersymmetry equations
 (\ref{covpure1},\ref{covpure2}) if and only if the fields \emph{before} the
 transformation satisfy the following equations, which can be
 regarded as a deformation of those in (\ref{covpure1},\ref{covpure2})
 determined by the fluxes associated with $P$.
 \bea  \Gamma^M \nabla_M(e^{2A - \phi} e^{B} \wedge \Phi_1) & = & 0 , \label{covpure1twist} \\
 \Gamma^M \nabla_M(e^{2A - \phi} e^{B} \wedge \Phi_2) & = & e^{2A-\phi} \Gamma^M \del_M A \wedge e^{B} \wedge \bar{\Phi}_2 \mp (-1)^n  \frac{i}{8} e^{3A}
\cK_6 e^{B} F.  \label{covpure2twist}
 \eea Here, $n$ is the number of $\Lambda_i$ factors that appear
 in the definition of $P = C_n S, S \in Spin^+(d,d)$.

Before we move on to the next subsection, we  would like note that
the transformation of pure spinor equations under a non-constant
$O(d,d)$ transformation was also studied in \cite{Andriot:2009fp}
and \cite{Andriot:2010ju}. They called such transformations twist
transformations and also used them  as solution generating
transformations in Type II theory.

 \subsection{Invariance under NATD}\label{s3}

As we discussed in section \ref{review}, the transformation under
NATD of the fields in the NS-NS sector can be performed via the
the action of the matrix $T_{{\rm NATD}}$ given in
(\ref{NATDmatrix}). Accordingly, the transformation of the RR
fields can be performed via the projected element $S_{{\rm NATD}}$
under the double covering homomorphism between $Pin(d,d)$ and
$O(d,d)$, see equations (\ref{NATDH})-(\ref{NATDd}). An important
point that should be stressed here is that $S_{{\rm NATD}}$ and
$T_{{\rm NATD}}$ act on the so-called \emph{untwisted fields}
$g(X), B(X), \phi(X),
 \Phi(X)$ and $F(X)$. These untwisted fields depend only on
 $10-dimG$ coordinates, where $dimG$ is the dimension of the
 non-Abelian isometry group $G$\footnote{For simplicity,  we assume here that the
 action of $G$ is free. However, the whole argument can be
 extended to the case where isotropy group of the action of  $G$
 is non-trivial, see \cite{Borsato}. } and is related to the
 background fields $g(X, \theta), B(X,\theta), \phi(X,\theta),
 \Phi(X,\theta)$ and $F(X,\theta)$ exactly as in
 (\ref{NATDH})-(\ref{NATDd}), where we replace the NATD
 coordinates $\nu$ with the space-time coordinates $\theta$ associated
 with the isometry directions\footnote{We choose a coordinate system adapted to the isometries so that the fields can be written
 in this separated form.} and the matrices $T_{{\rm NATD}}$
 and $S_{{\rm NATD}}$ with $L$ and $S_L$, respectively with, \be \label{geomtwist} L = \left(\begin{array}{cc}
                                  l^T & 0 \\
                                  0 & l^{-1}
                                  \end{array}\right), \ee and $S_L
                                  \in Spin^+(d,d)$ is such that
                                  $\rho(S_L) = L$.
 Here, $l$ is the $GL(10)$ matrix
obtained by embedding the  $GL(d)$ matrix $l_d$ with components
$(l_d)^I_{\ i} = l^I_{\ i}$ such  that $(l_d)^I_{\ m} = l^a_{\ i}
= 0$ and $(l_d)^a_{\ m} = \delta^a_{\ m}$. $l^I_{\ i}$ are
components of the left invariant 1-forms $\sigma^I = l^I_{\ i}
d\theta^{i}$ on $G$ defined from the Maurer-Cartan form: $g^{-1}
dg = \sigma^I T_I$ with $T_I$ forming a basis for the Lie algebra
${\cal{G}}$ of the isometry group $G$. For more details see
\cite{ayb2}. We also assume that the pure spinors associated with
the background respect the isometry so that (\ref{transphi}) also
holds for both pure spinors\footnote{Note that the second equality
in (\ref{transphiL}) is valid due to the special form of $L$ and
$S_L$, see equations (4.22)-(4.23) in \cite{ayb2}.} : \be
\label{transphiL} \Phi(X,\theta) = \sqrt{G} \ S_L(\theta) \ . \
\Phi(X) = \sqrt{{\rm det} \ l} \ e^{-B'(X,\theta)} S_L(\theta)
e^{B(X)} \Phi. \ee Now suppose that the background we start with
preserves at least ${\cal{N}}=1$ supersymmetry so that the pure
spinor equations (\ref{pure1},\ref{pure2}) are satisfied.
According to the discussions in the previous subsection and the
paragraph above, this means that the untwisted fields (which have
no dependence on the isometry directions) satisfy the deformed
pure spinor equations (\ref{covpure1twist},\ref{covpure2twist}),
where the deformation is determined by the flux associated with
the matrices $L$ and $S_L$. But this is just geometric flux with
$f_{ij}^{\ k} = C_{ij}^{\ k}$, see \cite{ayb2}. Now we act on
these untwisted fields with the NATD matrices
(\ref{NATDmatrix},\ref{NATDS}) as in (\ref{NATDH}-\ref{NATDd}) to
generate the NAT dual background. The resulting fields satisfy the
field equations of Type II supergravity as was shown in
\cite{ayb2}  by embedding these equations in DFT. To check
supersymmetry of the dual background we also transform the
untwisted pure spinors of the initial background (that is, the
pure spinors $\Phi(X)$ in (\ref{transphiL}) rather than
$\Phi(X,\theta)$) as in (\ref{transphi}) with $P = S_{{\rm
NATD}}$. Now we have to check whether these new pure spinors
$\Phi(X,\nu)$ still satisfy the supersymmetry equations
(\ref{pure1},\ref{pure2}). As discussed in the previous
subsection, this is equivalent to checking whether the untwisted
pure spinors satisfy the deformed supersymmetry equations
(\ref{covpure1twist},\ref{covpure2twist}), where the deformation
is determined by the flux associated with the NATD matrix $S_{{\rm
NATD}}$. As discussed above, due to the special form of the NATD
matrix: $S_{{\rm NATD}} = C_n S_{\theta}$, the associated flux can
be computed by calculating the flux associated with $S_{\theta}$
first (which gives the H-flux) and then raising one index with
$J^6_n$. This yields geometric flux with $f_{ij}^{\ k} = C_{ij}^{\
k}$, and we already know that the untwisted pure spinors satisfy
these deformed equations due to the existence of isometry
respected by the initial background and the pure spinors
associated with it. This completes the proof that a background
that preserves ${\cal{N}}=1$ supersymmetry will still be
supersymmetric after NATD.

\section{Examples}\label{examples}
In this section, we will demonstrate  how the NATD transformation
formulas (\ref{NATDH}-\ref{NATDd}) and (\ref{transphi}) work by
looking at a specific class of Type IIB backgrounds, which were
first studied in \cite{1509.04286}. The topology of the background
we will study is $R_{1,3}\times {\cal M}_3\times S^3$ so that
there is an $SU(2)$ isometry associated with $S^3$, which can be
utilized to perform NATD.

The ansatz for the metric and the 5-form flux  is
\bea\label{ansatz} ds^2 & = & e^{2A}dx^2_{1,3}+ ds^2({\cal M}_3)+
\sum_{i=1}^3 \big(e^{i}\big)^2,
\\ \nonumber  {\cal F}_5 & = & {\cal F}_2 \wedge e^1 \wedge e^2
\wedge e^3 \\ \nonumber F_5 & = & (1+ \star){\cal F}_5 = {\cal
F}_2 \wedge e^1 \wedge e^2 \wedge e^3 - e^{4A} \star_3 {\cal F}_2
\wedge Vol_4
  \eea and $ F_1 = F_3 = B = \phi = 0.$ ${\cal F}_2$ is a 2-form, $\star_3$ is the Hodge star operator on  ${\cal M}_3$, and $A$ is the warp
  factor. It is a function which has dependence only on the
  coordinates of ${\cal M}_3$. $S^3$ is assumed to be fibered over
  ${\cal M}_3$ and hence the vielbeins $e^i$ on $S^3$ have the form
\be\label{ei} e^i  =  \l_i (\s_i + {\cal A}_i). \ee Here, ${\cal
A}_i$ are 1-forms on ${\cal M }_3$ and $\l_i$ are functions on
${\cal M }_3$. The forms $\s^i$ are left invariant 1-forms for the
isometry group $SU(2)$ so that $d\s^i =\frac{1}{2} \epsilon^i_{jk}
\s^j \wedge \s^k$. We denote    the left invariant vector fields
$L_i$, so  $ \operatorname{i}_{L_i} \s^j =\delta_i^j$. We also
define (as in \cite{1509.04286}) a set of undetermined frame
fields $h^i$ so that \be ds^2({\cal M}_3)=\sum_{i=1}^3 (h^i)^2.
\label{hi} \ee

Another assumption that is made in \cite{1509.04286} is that  this
geometry preserves at least ${\cal{N}}=1$ supersymmetry in four
dimensions in the form of an $SU(3)$ structure  characterized by
the following 2-form $J$ and 3-form  $\Omega$  which are given by
means of  a vielbein $e^{i}$ and frame fields $h^i$:
\be\label{SU3structure}
  J = h^3 \wedge e^3+ e^1\wedge e^2 +  h^1 \wedge h^2 \ , \quad  \Omega = (h^3 + i e^3)\wedge (e^1 + i e^2 ) \wedge (h^1+ i h^2) \ .  \ee

As discussed in Appendix \ref{app1},  $SU(3)$ structure  can be
regarded as a special case of $SU(3) \times SU(3)$ structure with
associated pure spinors of the  form (\ref{genel3}). In our case,
setting  $\Phi_1=\Phi_-$ and $\Phi_2=\Phi_+$ we have
\be\label{purespinors}
  \Phi_+ = \frac{1}{8}  e^{i \th_+} e^{A} e^{-i J}  \ , \quad \Phi_- = -\frac{i}{8}  e^{i \th_-} e^A \Omega \ee

Due to assumption of preservation of supersymmetry, these pure
spinors must satisfy the equations (\ref{pure1},\ref{pure2}). As
shown in \cite{1509.04286}, this forces  $\th_+= \frac{\pi}{2}$
and ${\cal{A}}_1 = {\cal{A}}_2 = 0$. The possible values for
$\th_-$ for different geometries is given in Appendix B of
\cite{1509.04286}. Comparing (\ref{purespinors}) with
\eqref{genel3} one can see that it is of the general form of a
general $SU(3)$ pure spinor with $a = e^{i \theta_- / 2} e^{i
\theta_+ / 2} e^{A/2}$ and $b = e^{i \theta_- / 2} e^{-i \theta_+
/ 2} e^{A/2}$, which  satisfy $|a|^2=|b|^2=e^A$.

The ansatz \eqref{ansatz} is general enough to cover many examples
important for AdS/CFT duality, notably $AdS_5\times T^{1,1}$,
$AdS_5\times Y^{p,q}$ and $AdS_5\times S^5$. The detailed
description of how these backgrounds fall within this general
ansatz can be found in Appendix B of \cite{1509.04286}. For
example,  for $T^{1,1}$ background the required values are as
follows: $$A=\log r , \ \ \ {\cal A}_3=\cos\theta d\varphi ,\ \
\theta_{-}=0, $$ $$\l_1=\l_2=\frac{1}{\sqrt{6}}, \
\l_3=\frac{1}{3},\ \ \  h^1=\frac{1}{\sqrt{6}}\sin\theta d\varphi,
\ h^2=\frac{1}{\sqrt{6}}d\theta,\ h^3=\frac{dr}{r}.$$

On the other hand, the required values for the $AdS_5\times S^5$
background  are: $$ A=\log 2R,\ \ \ {\cal A}_3= 0, \ \
\theta_{-}=\beta,\ \l_1=\l_2=\l_3=\cos\alpha,$$ $$h^1=2 \frac{ R
\cos\alpha d\alpha +\sin\alpha dR}{R}, \ h^2= 2 \sin\alpha
d\beta,\ h^3=2\frac{\cos\alpha dR -  R \sin\alpha d\alpha}{R}. $$

Now, we perform the NATD transformation of the background
described by the ansatz \eqref{ansatz}. We begin with the
transformation of  the metric and the B-field. For this we use
\eqref{transE} where $T $ is obtained by embedding $ T_{{\rm
NATD}}$ in (\ref{NATDmatrix}) in $O(6,6)$ in the usual way. We
will call this $O(6,6)$ matrix also $ T_{{\rm
NATD}}$.\footnote{Note that, since the isometry group $SU(2)$ is
three dimensional, the matrix (\ref{NATDmatrix}) is in $O(3,3)$.}
Then we read off the transformed metric and the transformed
B-field from the symmetric and antisymmetric parts of $E'$,
respectively: \be \label{newQ} E'( g', B') = T_{{\rm NATD}} . \
E(g, B) \ee
 \be g' = \frac{E' + E'^t}{2}, \
\ B' = \frac{E' - E'^t}{2} \ee As mentioned before, this
transformation is equivalent to what is given in \eqref{NATDH}. We
refer to \cite{ayb2} for details. This gives \bea
\label{newgB}\nonumber d s'^2 & = & e^{2A} dx_{1,3}^2+
ds^2({\cal{M}}_3)+ \frac{1}{\Delta}\bigg( (\n_i \n_j +
\frac{\l_1^2 \l_2^2 \l_3^2}{\l_{(i)}^2} \delta_{(i)j})\  d\n_i \
d\n_j -2 \l_3^2 \l_2^2 \n_2 \  d\n_1\ {\cal A}_3 \\ & &+ 2 \l_3^2
\l_1^2 \n_1\ d\n_2\ {\cal A}_3 +(\l_3^2 \Delta-
4\l_3^4(\l_1^2\l_2^2+ \n_3^2) )\ {\cal A}_3 \ {\cal A}_3 \bigg)
\nonumber \\ B' & = & -
\frac{1}{\Delta}\bigg(\frac{1}{2}\epsilon_{ijk}\n_i\lambda_i^2 \
d\n_j\wedge d\n_k+\l_3^2 \n_3 \n_1 \ d\n_1 \wedge {\cal A}_3+
\l_3^2 \n_3 \n_2 \ d\n_2 \wedge {\cal A}_3 +(\l_3^2 \n_3^2+ \l_1^2
\l_2^2 \l_3^2) \ d\n_3 \wedge {\cal A}_3\bigg), \nonumber \\
\Delta & = & G^{-1} = \l_1^2\l_2^2\l_3^2+ \l_1^2 \n_1^2+ \l_2^2
\n_2^2+ \l_3^2 \n_3^2 \nonumber \\ e^{-2\phi'}& = &  \Delta  \eea
These are the same as the results obtained in \cite{1509.04286}
(except for a sign difference in the B-field, see footnote
\eqref{Bconvention}).

Next, we  perform the NATD transformation of the  RR flux  $F_5$
from the transformation rule \eqref{NATDS}, with $S_{{\rm NATD}}$
 \eqref{SNATDmatrix}. To this end, it is convenient to write the
spinor field $F$ that packages the RR fluxes   as a
non-homogeneous differential form as follows (see \cite{ozer,
ayb2}): \be F = \sum_p \left(F^{(p)} + F_i^{(p-1)} \sigma^i +
\frac{1}{2} F_{ij}^{(p-2)} \sigma^i \wedge \sigma^j + F^{(p-3)}
\sigma^1 \wedge \sigma^2 \wedge \sigma^3 \right), \ee where each
$p-$form is decomposed according to how many legs it has along the
$SU(2)$ directions. This non-homogeneous differential form maps to
a Clifford algebra element in the usual way where we identify the
element $\sigma^i$ with the Clifford algebra element $\psi^i$, for
$i= 1,2,3$, see \eqref{psi}. It has the following form: \be
 F = \sum_p \left(F^{(p)} + F_i^{(p-1)} \psi^{i} +
\frac{1}{2} F_{ij}^{(p-2)} \psi^{i}  \psi^{j} + F^{(p-3)} \psi^{1}
 \psi^{2}  \psi^{3} \right) \ee Then, the spinorial action of
$\psi^i$  on $F$ is given by wedge product, whereas the spinorial
action of $\psi_i$ is given  by contraction \cite{ozer, ayb2}: \be
\label{action} \psi^i. F = \psi^i \wedge F, \ \ \ \psi_i . F =
\operatorname{i}_{\psi_i}F. \ee

Since there is no B-field, we will first calculate  the action of
$C S_{\theta}$ on differential forms then apply $e^{ -B'}.$ The
action  of $S_{\theta}$ in \eqref{SNATDmatrix} on  a
non-homogeneous differential form $\a$ is as follows:
 \bea \label{Saction} S_{\theta}\cdot \a & = &  e^{-\theta} \wedge \a =
  \a + \nu_k \ \epsilon_{ij}^{\ \ k}\
 \psi^i \wedge \psi^j  \wedge \a
\\ \nonumber  & = &  \a+ \n_1\
\psi^2 \wedge \psi^3 \wedge \a+ \n_2 \ \psi^3 \wedge \psi^2 \wedge
\a + \n_3\ \psi^1 \wedge \psi^2 \wedge \a \eea

On the other hand, the action of $C$ given   in \eqref{chargeconj}
can be calculated by using \eqref{action}.  We calculate the
following NATD transformed RR flux.
\bea \nonumber F'_5& = & e^{- B'} C S_{\theta}\ F_5 \\ \nonumber &
= & \l_1 \l_2 \l_3\ {\cal F}_2 - \l_1 \l_2 \l_3\ B' \wedge  {\cal
F}_2  - \l_1 \l_2 \l_3\ \psi^3 \wedge {\cal A}_3 \wedge {\cal F}_2
+ e^{4A}
Vol_4  \wedge  \nu^i  \psi^i \wedge \star_3  {\cal F}_2 \\
& & - e^{4A} Vol_4  \wedge   \psi^1 \wedge   \psi^2 \wedge \psi^3
\wedge \star_3  {\cal F}_2  - B'  \wedge e^{4A} Vol_4 \wedge \nu^i
\psi^i \wedge \star_3  {\cal F}_2 \eea where the Hodge duality
$\star_3 $ is taken with respect to the transformed metric. This
polyform packages all the RR fluxes of the NAT dual background,
which we read off (after identifying $\psi^i$ with $d\nu_i$) to
be: \bea \label{fluxes}
 F'_2 & = & \l_1 \l_2\l_3 {\cal F}_2,\\
  F'_4 & = & (-B'+{\cal A}_3\wedge d\n_3)\wedge  F'_2,\\
  F'_6 & = & \star_{10} F_4' = e^{4A}Vol_4\wedge \n_i\ d\n_i\wedge\star_3{\cal F}_2,\\
  F'_8& = & -\star_{10} F_2' = -B' \wedge  F'_6+ e^{4A}Vol_4\wedge \star_3{\cal F}_2 \wedge dv_1\wedge d\n_2\wedge d\n_3.
\eea These agree with the results  obtained in \cite{1509.04286}
(up to sign differences in $B'$, and the 6- and 8-forms due to
differences
in conventions, see footnote \eqref{Bconvention}).

Finally, we will  apply the NATD transformation rule
\eqref{transphi} (with $P = S_{{\rm NATD}}$) to the $SU(3)$ pure
spinors given in \eqref{purespinors} (which are known to satisfy
the supersymmetry equations (\ref{pure1},\ref{pure2}) ) and obtain
the NAT-dual pure spinors $\Phi'_{+}$ and $\Phi'_{-}$. The
explicit form of the transformed pure spinors are presented in
Appendix B. In obtaining the results there, we first calculate
$S_{\theta} \Phi_-$:
\begin{eqnarray*}  S_{\theta} \Phi_-  & = &  \Phi_- + \n_1\
\psi^2 \wedge \psi^3 \wedge \Phi_-+ \n_2 \ \psi^3 \wedge \psi^2
\wedge \Phi_- + \n_3\ \psi^1 \wedge \psi^2 \wedge \Phi_-
\\ \nonumber & = & \Phi_- -\frac{i}{8}  e^{i \th_-} e^A\{ \l_1 \n_1\ \psi^2 \wedge \psi^3 \wedge h^3 \wedge \psi^1 \wedge h^1+
i \ \l_1 \n_1\ \psi^2 \wedge \psi^3 \wedge h^3 \wedge \psi^1
\wedge h^2
\\ \nonumber & & -\l_1 \l_3 \n_1 \ \psi^2 \wedge \psi^3 \wedge {\cal A}_3  \wedge \psi^1 \wedge h^2
+i \ \l_2 \n_2\ \psi^3 \wedge \psi^1 \wedge h^3 \wedge \psi^2
\wedge h^1
\\ \nonumber & &- \ \l_2 \n_2\ \psi^3 \wedge \psi^1 \wedge h^3 \wedge \psi^2 \wedge h^2
-i \ \l_2 \l_3 \n_2\ \psi^3 \wedge \psi^1 \wedge {\cal A}_3
\wedge \psi^2 \wedge h^2\}
\end{eqnarray*}
Applying ${\sqrt G} \ e^{-B'} C$ to $S_{\theta} \Phi_-$ above, we
obtain $\Phi'_{-}$, whose explicit form is given  in \eqref{new-}.

Now we calculate $S_{\theta} \Phi_+$:
\begin{eqnarray*}  S_{\theta} \Phi_+  & = &  \Phi_+ + \n_1\
\psi^2 \wedge \psi^3 \wedge \Phi_+ + \n_2 \ \psi^3 \wedge \psi^2
\wedge \Phi_+ + \n_3\ \psi^1 \wedge \psi^2 \wedge \Phi_+
\\ \nonumber & = & \Phi_+
+\frac{1}{8}  e^{i \th_+} e^{A} \{  \n_1\ \psi^2 \wedge \psi^3 -i
\ \l_3 \n_1  \ \psi^2 \wedge \psi^3 \wedge h^3 \wedge {\cal A}_3
-i\ \n_1 \ \psi^2 \wedge \psi^3 \wedge h^1 \wedge h^2
\\ \nonumber & & +  \n_2\ \psi^3 \wedge \psi^1
-i \ \l_3 \n_2  \ \psi^3 \wedge \psi^1 \wedge h^3 \wedge {\cal
A}_3 -i\ \n_2 \ \psi^3 \wedge \psi^1 \wedge h^1 \wedge h^2
\\ \nonumber & &+  \n_3\ \psi^1 \wedge \psi^2
-i \ \l_3 \n_3  \ \psi^1 \wedge \psi^2 \wedge h^3 \wedge {\cal
A}_3 -i\ \l_3 \n_3 \ \psi^1 \wedge \psi^2 \wedge h^3 \wedge \psi^3
\\ \nonumber & & -i\ \n_3 \ \psi^1 \wedge \psi^2 \wedge h^1 \wedge h^2
+\l_3 \n_3 \ \psi^1 \wedge \psi^2 \wedge \psi^3\wedge h^1 \wedge
h^2 \wedge h^3\}
\end{eqnarray*} Applying ${\sqrt G} \ e^{-B'} C$  to $S_{\theta} \Phi_+$ we obtain $\Phi'_{+}$, whose
explicit form is given in \eqref{new+}.

One can check by direct computation that the transformed pure
spinors $\Phi'_{-}$ and $\Phi'_{+}$ can be written  in the
following form: \bea \label{su2spinor-} \Phi'_{-} & = &
-\frac{i}{8} e^{A} e^{i \theta_-} e^{\frac{1}{2} z\wedge \bar z}
\wedge \omega
\\
\label{su2spinor+}
  \Phi'_{+} &=& - \frac{1}{8} e^{i \theta_+} e^{A} e^{-i j} \wedge z,
 \eea where the complex 1-form  $z=v+iw$, and the real and complex 2-forms  $j$  and
$\omega$ are as given below
\bea \label{z}
z & = &- \frac{1}{\sqrt \Delta}\bigg((\l_1 \l_2 \l_3+i \ \l_3 \n_3)\ h^3- \n_1\ d\n_1-  \n_2\ d\n_2-(\n_3-i \ \l_1 \l_2)\ d\n_3 \bigg)\\
\label{j} j & = & \frac{1}{\Delta}\bigg( \Delta \ h^1\wedge h^2+
\l_1 \l_2 \l_3^2 \ d\n_1 \wedge  d\n_2+
 \l_1 \l_2 \l_3^2 \n_1 \ d\n_1 \wedge {\cal A}_3+ \l_2^2 \l_3 \n_2 \ d\n_1 \wedge h^3 \\ \nonumber
   & &
 -\l_1 \l_2 \l_3^2 \n_2 \ {\cal A}_3 \wedge  d\n_2
+ \l_1^2 \l_3 \n_1 \  h^3\wedge  d\n_2 -( \l_2^2 \l_3 \n_2^2 +
\l_1^2 \l_3 \n_1^2)\ {\cal A}_3 \wedge  h^3 \bigg)\\ \nonumber
\label{omega}
\omega & = &  \frac{1}{\sqrt \Delta}\bigg( \l_2 \l_3 \ h^1\wedge
d\n_1+i\ \l_1 \l_3 \ h^1\wedge  d\n_2 +(\n_1 \l_1+i\ \n_2 \l_2)
h^1 \wedge h^3 +(i \ \n_1 \l_1- \n_2 \l_2) h^2 \wedge h^3
 \\
   & &+i \ \l_2 \l_3 \ h^2\wedge  d\n_1
-\l_1 \l_3 \ h^2\wedge  d\n_2 -(i\ \l_2 \l_3 \n_2 + \l_1 \l_3
\n_1)\  h^2 \wedge {\cal A}_3 \bigg). \eea

Comparing \eqref{su2spinor-},\eqref{su2spinor+} with
\eqref{genel2} one can see that they define an  $SU(2)$ structure,
as can be seen by taking $a = e^{i \theta_- / 2} e^{i \theta_+ /
2} e^{A/2}$ and $b = e^{-i \theta_- / 2} e^{i \theta_+ / 2}
e^{A/2}$ in \eqref{genel2}. Note that  $|a|^2=|b|^2=e^A$,  as
needed. So, under NATD, a background with $SU(3)$ structure is
transformed to a background with $SU(2)$ structure, as has been
demonstrated many times in the literature previously, in
particular in \cite{1509.04286,1305.7229}.

The results we present in (\ref{su2spinor-}-\ref{omega}) are in
 agreement with those obtained in \cite{1509.04286}\footnote{To be more precise, the results presented in in equation (5.4) of \cite{1509.04286}
    differ from our results in (\ref{su2spinor-},\ref{su2spinor+}) with an extra  -1  factor in $\Phi'_{+}$ and with a $-i$ factor in
  $\Phi'_{-}$ although the differential forms (5.6),(5.7) in \cite{1509.04286} and
  ours in \eqref{z}-\eqref{omega}  are exactly the same. However, we checked that the pure spinors (\ref{su2spinor-},\ref{su2spinor+}) satisfy the
  pure spinor equations \eqref{covpure1},\eqref{covpure2}.}.
 Whether these transformed pure spinors satisfy the supersymmetry equations
(\ref{pure1},\ref{pure2}) was checked in \cite{1509.04286} by
direct computation. The results we obtained in Section \ref{s3}
make such a calculation redundant. Indeed, the pure spinors
\eqref{su2spinor-},\eqref{su2spinor+} are obtained through the
action of $S_{{\rm NATD}}$ and we have proved that this
transformation maps solutions of (\ref{pure1},\ref{pure2}) to new
solutions.

\section{Conclusions and Outlook}\label{conclusion}

In this paper, we studied how the pure spinor equations
(\ref{pure1},\ref{pure2})  transform under NATD. These are
equations to be satisfied  for preservation of ${\cal{N}}=1$
supersymmetry in compactifications of Type II string theory to
four dimensions. Our approach in analyzing supersymmetry under
NATD is different from those in the literature in that we exploit
the recently discovered fact that NATD can be described as an
$O(d,d)/Pin(d,d)$ (in the NS-NS/RR sectors) transformation.
Although this is a coordinate dependent  transformation we start
in section \ref{s1} by considering constant $Pin(d,d)$
transformations. Writing the equations (\ref{pure1},\ref{pure2})
in terms of DFT fields makes it easy to show that they are
$Pin(d,d)$ covariant. This then means that solutions of these
equations will be mapped to new solutions under $Pin(d,d)$. This
analysis can be regarded as a generalization of those carried out
in \cite{0807.4527}  and \cite{0606257}, where the behavior of
pure spinor equations under Abelian T-duality (a certain type of
$O(1,1)/Pin(1,1)$ transformation) and LM deformations (a certain
type of $O(2,2)/Pin(2,2)$ transformation) was studied,
respectively.

Since the NATD matrix is coordinate-dependent, further analysis is
needed to see whether solutions are mapped to solutions under
NATD. This is done in section \ref{s2}. We show in that section
that this is indeed the case, due to the simple fact that the
fluxes generated by the NATD matrix (regarded as a twist matrix
within the formalism of GDFT) is the same as the geometric flux
associated with the isometry group that is used to perform NATD.
This idea of `preservation of flux' has been used before to
analyze field equations of supergravity under NATD in \cite{ayb2},
under YB deformations in
\cite{Borsato:2020bqo,Borsato2,Borsato3,Hassler:2020tvz,Borsato:2021vfy}
and under U-duality transformations in \cite{Gubarev}.

As we emphasized before, our approach here in analyzing
supersymmetry equations under NATD is novel, as it implements NATD
as an $O(d,d)/Pin(d,d)$  transformation. We believe that this
starting point is quite useful, as it has been realized in various
works recently that there are other interesting $O(d,d)$
transformations that can be utilized to generate new supergravity
backgrounds, notably related with integrable deformations of
string sigma models \cite{Lust},\cite{Borsato2, Borsato3},
\cite{Sakamoto:2017cpu}-\cite{Codina:2020yma}. The approach taken
here would also be useful to analyze supersymmetry of such
backgrounds. Also, viewing NATD as a $Pin(d,d)$ transformation
makes it easier to apply it to other backgrounds, which fall
outside the ansatz considered in section \ref{examples} with
different isometry groups and supporting a generic $SU(3) \times
SU(3)$ structure \cite{0609124},\cite{Koerber:2007hd},
\cite{Andriot:2008va}. The methods we employed here are well
suited to analyze the supersymmetry and structure group of the
resulting backgrounds. We plan to consider these issues in future
work.

\appendix
\section*{Appendices}

\section{$SU(3)$ and  $SU(2)$ Structures and Pure Spinors}
\label{app1}

The structure group of the  generalized tangent bundle $TM
\oplus T^*M$ of the six dimensional internal manifold $M$ reduces to $SU(3) \times SU(3)$
if there exists two globally defined $SU(3) \times SU(3)$ pure
spinors $\Phi_1$ and $\Phi_2$ of non-vanishing norm,
\cite{2005,0609124}.
 Adopting the conventions of \cite{0606257}, the explicit form of these pure spinors can be given as below:
\bea \label{SU(3)xSU(3)} && \Phi_+  = \frac{1}{8} \Big[ c_1 \bar
c_3 e^{- i j} +  c_2 \bar c_4 e^{i j} - i ( c_1 \bar{c}_4 \omega +
\bar{c}_3 c_2 \bar{\omega} ) \Big]\wedge e^{z \bar{z}/2} \, , \\
\label{A2} && \Phi_{-}  = \frac{1}{8} \Big[ i (c_2 c_4
\bar{\omega} - c_1 c_3 \omega) + ( c_2 c_3 e^{i j} - c_1 c_4 e^{-
i j}) \Big]\wedge z\, . \eea where $c_1,c_2,c_3,c_4$  are complex
functions on $M$. For a background of the form \eqref{metric}
requirement of existence of supersymmetric branes imposes that
$|c_1|^2+|c_2|^2=|c_3|^2+|c_4|^2 = e^{A}$
\cite{0606257}.\footnote{In fact only then the pure spinor
equations whose general form is given in equation (2.17) of
\cite{0606257} reduce to those in (\ref{pure1},\ref{pure2}). See
section 2 of \cite{0606257} for more details.} Here $z=v+iw$ is a
complex 1-form, $j$ is a real 2-form and $\omega$
 is a complex 2-form.

Reduction of the structure group of the tangent bundle $TM$ to
$SU(3)$ is equivalent to existence on $M$ of  an invariant real
2-form $J$ and a complex 3-form $\Omega$ satisfying the following
compatibility
conditions,\cite{salamon,Gauntlett:2003cy,DallAgata:2003txk,DallAgata:2004amo}.
\begin{equation}
\frac{i}{8}\Omega\wedge\bar{\Omega}=\frac{1}{3!}J\wedge J\wedge J,
\qquad J\wedge \Omega=0.\label{eq:SE_algebraic}
\end{equation}

$SU(3)$ structure can be regarded as a special case of $SU(3)
\times SU(3)$ structure \cite{hitchin,gualtieri} and the form of
the pure spinor describing the the $SU(3)$ structure is a special
case of \eqref{SU(3)xSU(3)} and \eqref{A2}
\cite{2005,0609124,0606257}
\begin{equation}\label{genel3}
{\rm SU(3):} \qquad \Phi_-= -i\ \frac{a  b}{8} {\Omega} \, ,
\qquad \Phi_+= \frac{a \bar b}{8} e^{- i J}
\end{equation}

Comparing to \eqref{SU(3)xSU(3)} and \eqref{A2}  we have $ J =j +
v\wedge w\ , \ \Omega  =  \omega\wedge(v + i w)$ and $c_1=a$,\
$c_3=b$, $c_2=c_4=0$. Due to the condition
$|c_1|^2+|c_2|^2=|c_3|^2+|c_4|^2 = e^{A}$ we need $|a|^2=|b|^2 =
e^A$. In \eqref{purespinors} we had $a = e^{i \theta_- / 2} e^{i
\theta_+ / 2} e^{A/2}$ and $b = e^{i \theta_- / 2} e^{-i \theta_+
/ 2} e^{A/2}$.

On the other hand, $SU(2)$ structure on $M$  is characterized by
the existence of a complex 1-form $z=v+iw$, a real 2-form $j$  and
a complex 2-form $\omega$ satisfying the following compatibility
conditions
\cite{salamon,Gauntlett:2003cy,DallAgata:2003txk,DallAgata:2004amo}:
\begin{eqnarray} \label{SU2SE}
\omega \wedge j  & = & 0, \\ \nonumber
 \operatorname{i}_{z} j  =  \operatorname{i}_{z} \omega & = &  0,\\ \nonumber
\omega \wedge \bar{\omega}& = &2 j \wedge j.
\end{eqnarray}

Again,  $SU(2)$ structure  can be regarded as a special case of
$SU(3) \times SU(3)$ structure \cite{hitchin,gualtieri} and the
form of the pure spinor describing the $SU(2)$ structure is a
special case of \eqref{SU(3)xSU(3)} and \eqref{A2}. The
corresponding pure spinors are \cite{2005,0609124,0606257}:
\begin{equation}
  \label{genel2}
{\rm SU(2):} \qquad  \Phi_-=-\frac{ab}{8} e^{-i\,j}\wedge(v+i w) \
, \qquad \Phi_+= -i\frac{a\bar b}{8}   e^{-i\,v\wedge w}
\wedge\omega.
\end{equation}
Comparing to \eqref{SU(3)xSU(3)} and \eqref{A2}  we have $c_2= c_3
= 0$ and $c_1 = a$, $c_4 = b$, again with $|a|^2=|b|^2 = e^A$ . In
\eqref{su2spinor-}, \eqref{su2spinor+} we had $a = e^{i \theta_- /
2} e^{i \theta_+ / 2} e^{A/2}$ and $b = e^{-i \theta_- / 2} e^{i
\theta_+ / 2} e^{A/2}$.
\section{NAT-dual Pure Spinors}

Transformation of the $SU(3)$ pure spinors given in
\eqref{purespinors} under the NATD transformation yields the pure
spinors presented  below:
\begin{eqnarray} \label{new+}
 \nonumber  \Phi'_+ & = & -\frac{1}{8\sqrt \Delta}  e^{i \th_+} e^A \bigg( \n_1 \ d\n_1+ \n_2\  d\n_2
 + (\n_3-i \l_1 \l_2) \ d\n_3-(\l_1 \l_2 \l_3+i \l_3 \n_3)\ h^3
  \\ \nonumber & &
  -( \ \l_1^2\l_2^2\l_3^2+i\  \l_1 \l_2 \l_3^2 \n_3)\ d\n_1 \wedge d\n_2 \wedge d\n_3
  +(i\ \l_1 \l_2 \l_3- \l_3 \n_3)\ h^1\wedge h^2 \wedge h^3
  \\ \nonumber & &-i\ \n_1 \ h^1\wedge h^2 \wedge d\n_1
   -i\ \n_2 \ h^1\wedge h^2 \wedge d\n_2
   -(\l_1 \l_2+i\ \n_3)\ h^1\wedge h^2 \wedge d\n_3
   \\ \nonumber & &
   -\frac{1}{\Delta}(i\ \l_3^3 \n_3^2+\l_1 \l_2 \l_3^3 \n_3-i \ \l_3 \Delta)\ d\n_1 \wedge d\n_2 \wedge h^3
  \\ \nonumber & & -\frac{1}{\Delta}(i \ \l_2^2 \l_3 \n_2 \n_3 +\l_1 \l_2^3 \l_3 \n_2)\ d\n_1 \wedge h^3\wedge d\n_3
   \\ \nonumber & & -\frac{1}{\Delta}(i \ \l_1^2 \l_3 \n_1 \n_3 +\l_1^3 \l_2 \l_3 \n_1)\ h^3 \wedge d\n_2 \wedge d\n_3
   \\ \nonumber & &+\frac{1}{\Delta}(i \ \l_1 \l_2 \l_3^2 \n_1 \n_3 +\l_1^2 \l_2^2 \l_3^2 \n_1)\ d\n_1 \wedge d\n_3 \wedge {\cal A}_3
   \\ \nonumber & & +\frac{1}{\Delta}(i \ \l_1 \l_2 \l_3^2 \n_2 \n_3 +\l_1^2 \l_2^2 \l_3^2 \n_2)\ d\n_2 \wedge d\n_3 \wedge {\cal A}_3
   \\ \nonumber & & +\frac{1}{\Delta}( \l_1 \l_2 \l_3^2 \n_1 \n_3-i \ \l_1^2 \l_2^2 \l_3^3 \n_1 -i \ \l_2^2 \l_3 \n_1 \n_2^2  -i \ \l_1^2  \l_3 \n_1^3)\ d\n_1 \wedge h^3 \wedge {\cal A}_3
   \\ \nonumber& &+\frac{1}{\Delta}(i \ \l_1^2 \l_2^2 \l_3^3 \n_2-\l_1 \l_2 \l_3^3 \n_2 \n_3-i \ \l_2^2  \l_3 \n_2^3
   -\l_1^2 \l_3 \n_2 \n_1^2)\ d\n_2 \wedge h^3 \wedge {\cal A}_3
   \\ \nonumber & &-\frac{1}{\Delta}( \l_1 \l_2^3 \l_3 \n_2^2+\l_3 \l_2 \l_1^3 \n_1^2+i \ \l_2^2  \l_3 \n_2^2 \n_3
   +i \ \l_1^2 \l_3 \n_3 \n_1^2)\ d\n_3 \wedge h^3 \wedge {\cal A}_3
   \\ \nonumber & &
   +\frac{1}{\Delta}( i \ \l_1 \l_2 \l_3^3 \n_3- \l_3^3 \n_3^2+ \l_3 \Delta)\ h^1\wedge h^2\wedge h^3\wedge d\n_1 \wedge d\n_2
 \\ \nonumber & & +\frac{1}{\Delta}(  \l_2^2 \l_3 \n_2 \n_3-i \ \l_1 \l_2^3 \n_2 \n_3) \ h^1\wedge h^2\wedge h^3\wedge d\n_1 \wedge d\n_3
    \\ \nonumber & & +\frac{1}{\Delta}( i\  \l_1^3 \l_2 \l_3 \n_1-  \l_1^2 \l_3 \n_1 \n_3)\ h^1\wedge h^2\wedge h^3\wedge d\n_2 \wedge d\n_3
\\ & & +\frac{1}{\Delta}(i \ \l_1^2 \l_2^2 \l_3^2- \l_1 \l_2 \l_3^2 \n_3)\ h^1\wedge h^2\wedge d\n_1 \wedge d\n_2 \wedge d\n_3
   \bigg)
\end{eqnarray}

Similarly, we transform $\Phi_-$ under NATD and obtain
\begin{eqnarray}\label{new-}
\nonumber  \Phi'_- & = &
    \frac{-i}{8 \sqrt \Delta}  e^{i \th_-} e^A \bigg( -\l_2 \l_3  \ d\n_1 \wedge h^1
  -i \ \l_1 \l_3  \ d\n_2 \wedge h^1+ (\n_1 \l_1+i\ \n_2 \l_2)\ h^1 \wedge h^3+i \ \l_2 \l_3\ h^2 \wedge d\n_1
 \\ \nonumber & & -\l_1 \l_3\ h^2 \wedge d\n_2- (i \ \n_1 \l_1-\ \n_2 \l_2)\ h^2 \wedge h^3 -(i \ \l_2 \l_3 \n_2+ \l_1 \l_3 \n_1)\ h^2 \wedge {\cal A}_3
 \\ \nonumber & & +\frac{1}{\Delta}(\l_1 \l_3^2 \n_1 \n_3+i \ \l_2 \l_3^2 \n_2 \n_3)\ h^3\wedge d\n_1\wedge h^1\wedge d\n_2
 \\ \nonumber & & +\frac{1}{\Delta}(i\ \l_1 \l_3^2 \n_1 \n_3- \ \l_2 \l_3^2 \n_2 \n_3)\ h^3\wedge d\n_1\wedge h^2\wedge d\n_2
 \\ \nonumber & & +\frac{1}{\Delta}(i\ \l_2^3  \n_2^2+ \l_1 \l_2^2 \n_2 \n_1-i\ \l_2 \Delta  )\ h^3\wedge d\n_3\wedge h^1\wedge d\n_1
\\ \nonumber& &+\frac{1}{\Delta}(\l_1 \Delta -i\ \l_1 \l_2 \n_2 \n_1- \l_1^3 \n_1^2 )\ h^3\wedge d\n_3\wedge h^1\wedge d\n_2
\\ \nonumber& &+\frac{1}{\Delta}(\l_2 \Delta +i\ \l_1 \l_2^2 \n_2 \n_1- \l_2^3 \n_2^2 )\ h^3\wedge d\n_3\wedge h^2\wedge d\n_1
\\ \nonumber & &
+\frac{1}{\Delta}(i \ \l_1 \Delta + \l_2 \l_1^2 \n_2 \n_1- i\
\l_1^3 \n_1^2 )\ h^3\wedge d\n_3\wedge h^2\wedge d\n_2
\\ \nonumber & &-\frac{1}{\Delta} \l_1 \l_2 \l_3(\l_1 \n_1+i \ \l_2 \n_2)\ d\n_1 \wedge d\n_2 \wedge d\n_3 \wedge h^1
\\ \nonumber& &
+ \frac{1}{\Delta}\l_1 \l_2 \l_3(\l_2 \n_2-i \ \l_1 \n_1) \ d\n_1
\wedge d\n_2 \wedge d\n_3 \wedge h^2
\\ \nonumber& &+\frac{1}{\Delta}(i \ \l_1 \l_3^2 \n_1 \n_2 \n_3- \l_2\l_3^2 \n_2^2 \n_3)\ h^3\wedge d\n_2 \wedge  h^2 \wedge {\cal A}_3
\\ \nonumber& &
+\frac{1}{\Delta}(i \ \l_1 \l_3^2 \n_1^2 \n_3- \l_2 \l_3^2 \n_1
\n_2 \n_3 )\ h^3\wedge d\n_1 \wedge  h^2 \wedge {\cal A}_3
\\ \nonumber & &
-\frac{1}{\Delta}(\l_1^2 \l_2^3 \l_3^2 \n_2+\l_2 \l_3^2 \n_2
\n_3^2-i\ \l_2^2 \l_1^3 \l_3^2 \n_1-i \ \l_1 \l_3^2 \n_1 \n_3^2)\
h^3\wedge d\n_3 \wedge  h^2 \wedge {\cal A}_3
\\ \nonumber& &
+\frac{1}{\Delta}( \l_1 \l_2^2 \l_3 \n_1 \n_2-i\ \l_1^2 \l_2  \l_3
\n_1^2 )\ d\n_1\wedge d\n_3 \wedge  h^2 \wedge {\cal A}_3
\\ & &
+\frac{i}{\Delta} ( \l_1 \l_2^2 \l_3 \n_2^2-\ \l_1^2 \l_2  \l_3
\n_1 \n_2 )\ d\n_2 \wedge d\n_3 \wedge  h^2 \wedge {\cal A}_3
\bigg)
\end{eqnarray}
Note that after the transformation $\psi^i$ are identified with
$d\n_i$. It can be checked by direct computation that the spinors
$\Phi'_+$ and $\Phi'_- $ can be written as in
\eqref{su2spinor-}-\eqref{omega}.

\section{Mukai pairing}
\label{appC}

Mukai pairing is the natural inner product on the Clifford module
$\wedge^\bullet T^*$ and described as follows.

$\langle \ , \ \rangle : S \otimes S \rightarrow \wedge^n T^*$:
\be \label{mukai} \langle \chi_1, \chi_2 \rangle = (\tau(\chi_1)
\wedge \chi_2 )_{{\rm top}} = \sum_j (-1)^j (\chi_1^{2j} \wedge
\chi_2^{n-2j} + \chi_1^{2j+1} \wedge \chi_2^{n-2j-1}), \ \ \
\chi_1, \chi_2 \in \wedge^\bullet T^*, \ee here $( )_{{\rm top}}$
denotes the top degree component of the form and the superscript
$k$ denotes the $k$-form component of the form. This is equivalent
to \be <\chi_1, \chi_2> = (\chi_1 \wedge
\lambda(\chi_2))_{{\rm{top}}}, \ee where $\lambda$ is the natural
linear extension of $\lambda$ in (\ref{lambda}) to a
non-homogeneous differential form.

Mukai pairing  is symmetric in dimensions $n \equiv 0, 1 $ (mod 4)
and is skew-symmetric otherwise:
 \be \label{mukaisymmetric} \langle \chi_1 , \chi_2
\rangle = (-1)^{n(n-1)/2} \langle \chi_2 , \chi_1 \rangle. \ee See
\cite{ozer} for details.

Mukai pairing has an important property related to the action of
the Spin group, \cite{gualtieri}: \be \label{mukaiinvariant}
\langle S \chi_1, S \chi_2 \rangle = \pm \langle \chi_1 , \chi_2
\rangle, \ \ S \in Spin(d,d).\ee

This follows from \be \label{spininv} \langle  P \chi_1, P \chi_2
\rangle= (P, P)\langle \chi_1, \chi_2 \rangle, \ee
 where $(,)$ is the natural indefinite inner product defined as
$$(X+\xi,X+\xi)=i_X\xi= \xi(X), \ \  X + \xi \in  T\oplus T^*.$$ Since $(P, P)=\pm 1$, when $P \in Spin(d,d)$,
\eqref{spininv} implies \eqref{mukaiinvariant}. In the special
case when $S \in Spin^+(d,d)$ we have $(S,S) = +1$, so Mukai
pairing is invariant under the connected component to identity,
$\operatorname{Spin}^{+}(d,d)$. See \cite{gualtieri} for further
details.

The NATD matrix is not an element of $Spin^+(d,d)$. However, due
its special form given in \eqref{SNATDmatrix} we still have \be
\label{mukainatd} <S_{NATD} \chi_1, S_{NATD} \chi_2 > = -<\chi_1,
\chi_2>. \ee This can be seen as follows: As discussed in detail
in \cite{dftRR,ozer} the  matrix that appears in the definition of
\eqref{SNATDmatrix} (and also of $\cK$ with $n=d$) is \be
\label{chargeconj} C_n = \Lambda_1 \cdots \Lambda_n, \ee where \be
\label{lambda2} \Lambda_i = (\psi^i - \psi_i). \ee Here, $\psi_i,
\psi^i$ are elements of the Clifford algebra $Cliff(d,d)$ given in
(\ref{psi}), and hence obey the commutation relations following
from (\ref{clifford}). Since $(\Lambda_i , \Lambda_i) =
-i_{\psi_i} \psi^i = -1$, repeated use of \eqref{spininv} gives
  \be
  <C_n \chi_1, C_n \chi_2> = (-1)^n <\chi_1, \chi_2>
  \ee
From this it follows (again using \eqref{spininv} repeatedly) \be
\label{mukainatd} <S_{NATD} \chi_1, S_{NATD} \chi_2 > = <S_\beta
C_3 \chi_1, S_\beta C_3 \chi_2> = <C_3 \chi_1, C_3 \chi_2> =
-<\chi_1, \chi_2>, \ee as claimed. Note that in the second
equality we used the fact that $S_\beta \in Spin^+(6,6)$.

In addition to the elements $\Lambda_i$ defined in
(\ref{lambda2}), it is also useful to define the elements \be
\label{lambdaplus} \Lambda^+_i = (\psi^i + \psi_i), \ee and \be
\label{chargeconj} C^+_n = \Lambda^+_1 \cdots \Lambda^+_n. \ee
From the Clifford commutation relations (\ref{clifford}) one can
easily compute  \be \label{anticommplus}
 \Lambda^+_i . \Gamma^M . (\Lambda^+_i)^{-1}
                    =\left\{
  \begin{array}{l l l}
   \Gamma_i & \quad \text{if\; $\Gamma^M = \Gamma^i$}\\
   \Gamma^i & \quad \text{if\, $\Gamma^M = \Gamma_i$\,.}\\
   - \Gamma^M & \quad \text{otherwise}
  \end{array} \right.
                    \ee
This then means that  $\rho(\Lambda^+_i) = h^+_i,$ where \be h^+_i = - \left(\ba{cc} 1 - E_i &  -E_i \\
                                     -E_i  & 1 - E_i \ea \right), \ \ (E_i)_{jk} = \delta_{ij}\delta_{ik}. \ee

The charge conjugation matrix which appears in the definition of
$\cK \in Spin(d,d)$ is $C_d$ for even $d$, whereas it is $C^+_d$
for odd $d$, as explained in \cite{dftRR}.

\section*{Acknowledgments}
This work is  supported by the Turkish Council of Research and
Technology (T\"{U}B\.{I}TAK) through the ARDEB 1001 project with
grant number 121F123.

\end{document}